# Liquid crystal-enabled electrophoresis of spheres in a nematic medium with negative dielectric anisotropy

Israel Lazo and Oleg D. Lavrentovich[†]

*Liquid Crystal Institute and Chemical Physics Interdisciplinary Program,*
*Kent State University, Kent, OH 44242*

We describe electrophoresis of spherical dielectric particles in a uniformly aligned nematic medium with a negative dielectric anisotropy. A spherical particle that orients the liquid crystal (LC) perpendicularly to its surface moves under the application of the uniform direct current (DC) or alternating current (AC) electric field. The electric field causes no distortions of the LC director far away from the sphere. Electrophoresis in the nematic LC shows two types of nonlinearity in the velocity vs. field dependence. The velocity component parallel to the applied electric field grows linearly with the field, but when the field is high enough, it also shows a cubic dependence. The most interesting is the second type of nonlinear electrophoresis that causes the sphere to move perpendicularly to the applied field. This perpendicular component of velocity is proportional to the square of the field. The effect exists only in a LC and disappears when the material is melted into an isotropic fluid. The quadratic effect is caused by the dipolar symmetry of director distortions around the sphere and is classified as an LC-enabled electrophoresis (LCEEP). The nonlinear electrophoretic mobility of particles in LCEEP offers a rich variety of control parameters to design 3D trajectories of particles for microfluidic and optofluidic applications.

*Keywords: Nonlinear electroporesis, Liquid Crystals, Colloids.*

## I. INTRODUCTION

Controlled manipulation and motion of small particles has become a topic of great interest over the last decade. Drug delivery, macromolecule separation, display of information and colloidal assembly are just a few examples of potential applications. Control over the particles can be achieved in a variety of ways, including the design of self-propelled particles. Historically, however, the most popular driving agent of particle motion has been the electric field. The most known technique is electrophoresis in which the charged particles are moved by an applied DC electric field. There is a growing interest in finding mechanisms for particle manipulation that uses an AC electric field, as with the latter, it is much easier to produce steady flows and to avoid undesirable electrochemical reactions. Clearly, the AC driving mechanism should somehow break the classic linear Smoluchowski relationship between the velocity of particles and the electric field **E** [1–7]. One of the interesting developments was a discovery that a broken symmetry of the particle can result in a nonlinear electrophoresis, with a quadratic dependence of the velocity on the field. The phenomenon was described for particles moving in an isotropic fluid, first by Murtzovkin [8] and then recently by Squires and Bazant [9] who introduced the term induced charged electrophoresis (ICEP). Experimentally,

ICEP was discovered by Velevs group for Janus spheres (half dielectric, half metallic) moving in water [10]. Another interesting aspect of the electrically driven motion is whether the velocity of the transported particle can be altered from a trajectory that is determined by the direction of the field and polarity of the particles charge. A particular case is the reversal of the electroosmotic flow of isotropic fluid observed when the frequency of the applied AC field changes [7, 11, 12]; the phenomenon is not fully understood [13, 14]. Electrically controlled motion of particles in an anisotropic fluid (better known as a liquid crystal, LC) is clearly more complicated than in the isotropic fluid. A replacement of an isotropic fluid with a LC should bring first of all some anisotropy in the velocity of particles transport parallel and perpendicular to the director **n̂** (the average direction of molecular orientation), associated with the different Stokes drag in these two directions, see the review by Stark et al. [15]. There are, however, some qualitative differences. Jákli et al [16, 17] described both rotational and translational motion of colloidal particles in LCs, attributing the rotational aspects to Quincke mechanism and showing a conversion of this rotation into a translational motion as a result of hydrodynamic interaction with the walls. Dierking et al [18] has reported on electromigration of microspheres in nematic LCs and noticed that the particles move under the AC field in the direction perpendicular to **E** and parallel to **n̂**; the velocity was measured to have a linear dependency on the field. A classic quintessential LC effect is the so-called backflow, i.e. generation of LC flow by director reorientations triggered by an electric field applied to a LC with dielectric anisotropy. The backflow effect was shown to be effective for transport and assembly [19, 20] of colloids in a LC; the prime driving force of transport is the field-LC interaction rather than the field-particle interaction. Ryzhkova, Podgornov and Haase [3] have reported on the classic electrophoretic experiment staged in a LC in which the electric field was acting on charged particles and found a nonlinear (cubic) term in the dependence of $\mathbf{v}$ on $\mathbf{E}$, in addition to the classic linear term. Finally, our group reported that the electrophoretic motion of a particle in an aligned LC has a pronounced nonlinearity with a quadratic dependence of $\mathbf{v}$ on $\mathbf{E}$ [21]. The electrophoretic motion was observed for particles that are spherical (an effect that is impossible in an isotropic fluid) and for particles that have no electric charge. This liquid crystal-enabled electrophoresis (LCEEP) was attributed to the dipolar character of director distortions around the particle. The asymmetry of the host in LCEEP plays a role similar to the asymmetry of particles themselves in ICEP described by Velev et al. for Janus spheres moving in water [10]. In our previous work on LCEEP [21], we focused mostly on the LC with a positive dielectric anisotropy $\epsilon_a = \epsilon_\parallel - \epsilon_\perp > 0$, where $\epsilon_\parallel$ and $\epsilon_\perp$ are the dielectric constants measured parallel and perpendicular to $\mathbf{E}$, respectively. The electrophoretic motion was observed when the electric field was applied parallel to the overall director $\hat{\mathbf{n}}_0$ ($\hat{\mathbf{n}}_0$ is fixed in space by surface alignment) so that $\hat{\mathbf{n}}_0$ is not influenced far away from the particle, to avoid the backflow-induced transport [19, 20]. In this paper, we expand the studies to the case $\epsilon_a < 0$. This means that the electric field does not cause any reorientation as long as $\mathbf{E} \perp \hat{\mathbf{n}}_0$. Such a LC allows us to use two mutually perpendicular field directions that do not perturb $\hat{\mathbf{n}}_0$ far from the particle: one perpendicular to the sandwich-type cell, and another one parallel to it, Fig.1. We demonstrate that the electric field causes two qualitatively different nonlinear electrophoretic effects in the LC. The first effect is a LCEEP that drives the spheres along the director $\hat{\mathbf{n}}_0$, Fig.1, with the velocity growing as $v_{LCEEP} \propto E^2$. This effect vanishes when the LC is melted into the isotropic phase by increasing the temperature of sample. The second effect is reminiscent of a classic nonlinear electrophoresis in the isotropic media [22, 23] with the velocity growing as $v_{13} \propto E + \alpha E^3$, where $\alpha$ is a non-vanishing nonlinearity coefficient; the nonlinear behavior of velocity is enhanced at higher fields and for larger particles.

Besides the difference in the even vs. odd dependency of $v$ on $E$, there is also an important difference in the direction of particle motion: the LCEEP velocity is perpendicular to the applied field, $\mathbf{v}_{LCEEP} \perp \mathbf{E}$ (and parallel to $\hat{\mathbf{n}}_0$), while the *standard* nonlinear electrophoresis velocity is parallel to the field, $\mathbf{v}_{13} \parallel \mathbf{E}$ so that $\mathbf{v}_{LCEEP} \perp \mathbf{v}_{13}$.

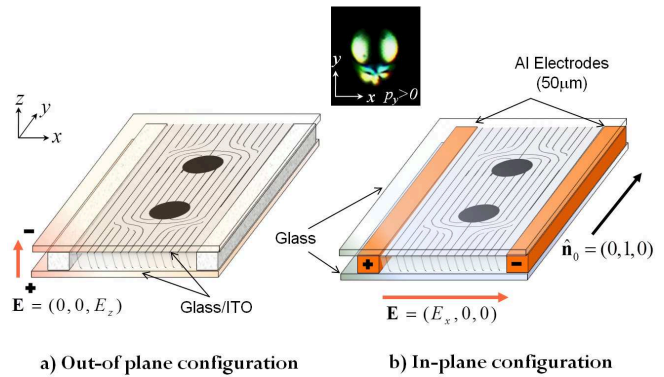

Figure 1: Scheme of electrophoretic experiment for spheres showing opposite polarities of $p_y$ in a LC with $\epsilon_a < 0$ aligned along $\hat{\mathbf{n}}_0 = (0, 1, 0)$. (a) Out-of plane configuration with the electric field $\mathbf{E} = (0, 0, E_z)$; (b) In-plane configuration with the electric field $\mathbf{E} = (E_x, 0, 0)$. The inset shows a polarizing microscope texture of a spherical particle with a hyperbolic hedgehog near its bottom.

## II. MATERIALS AND METHODS

We used N, N-didecyl-N-methyl (3-trimethoxysilyl-propyl) ammonium chloride (DDMAC) to induce perpendicular alignment of the director $\hat{\mathbf{n}}$ at the particles surface. To modify the surface, we dispersed some quantity of spheres in a solution of DDMAC/Ethanol $< 1$ wt% by sonication during 30 min. The excess of solvent is evaporated and the dry-treated spheres are mixed in the LC by ultrasonication during 10 min. Polarizing optical microscopy textures show that DDMAC yields a normal (homeotropic) alignment of LC at the surface of spheres. We used dielectric particles of various diameters, namely, silica $SiO_2$ spheres of $2a = (5.08 \pm 0.5)\mu$m (Bangs Laboratories), borosilicate glass spheres $2a = (9.6 \pm 1)\mu$m and $2a = (17.3 \pm 1.4)\mu$m (Duke Scientific). They were added to the nematic host in small quantities (less than 1 wt%) to avoid aggregation. We have chosen the room-temperature nematic mixture MLC7026-000 from Merck ($t_{ni} = 80$ °C ) with a negative dielectric anisotropy $\epsilon_a = -3.7$ (measured at $T = 25$ °C and 1 kHz). The samples represent a thin nematic layer (thickness $h$ varied from $10\mu$m to $60\mu$)m sandwiched between two glass plates. The glass substrates are coated with polyimide PI2555 (Microsystems) which is mechanically buffed to produce a uniform alignment along the rubbing direction in the plane of the cell, $\hat{\mathbf{n}}_0 = (0, 1, 0) = const$, Fig.1. The buffing procedure typically results in a small, 1°-2° *pretilt* angle between $\hat{\mathbf{n}}_0$ and the plate. To mitigate the possible role of the pretilt in the dynamic phenomena, the two plates were assembled with antiparallel rubbing directions, which means that $\hat{\mathbf{n}}_0$ is uniform but slightly tilted. The local radial configuration $\hat{\mathbf{n}}(\mathbf{r}) = \hat{\mathbf{r}}$ at the immediate vicinity of the particle needs to match the uniform director $\hat{\mathbf{n}}_0 = const$ away from the sphere. For the studied particles, the matching is achieved by a topolo-gical point defect, the so-called *hyperbolic hedgehog* located at short distance from the sphere [24], Fig. 1. There is an important symmetry breaking associated with the hyperbolic hedgehog, as the director distortions are polar, Fig. 1. We label the corresponding elastic dipole as **p** and direct it from the hedgehog (negative topological charge -1) towards the sphere (positive topological charge 1). The elastic dipole $\mathbf{p} = (0, p_y, 0)$ is parallel to $\hat{\mathbf{n}}_0 = (0, 1, 0)$ and can adopt two orientations, $p_y > 0$ or $p_y < 0$. In a typical sample, the particles with $p_y > 0$ and $p_y < 0$ are met with equal probability. Once chosen, the sign of elastic dipole cannot change as that would require a motion of the hyperbolic hedgehog around the sphere, or its splitting into a macroscopic ring, or melting the nematic sample into the isotropic phase, with all these processes characterized by a huge energy barrier. Under a polarizing optical microscope, the silica spheres appear as bright circles with a *Malthese* cross and a small *tail* associated with the hedgehog, Fig. 1 (inset). The Malthese cross results from the radial director near the sphere. As shown in Fig. 1, the electric field can be applied in two geometries with $\mathbf{E} \perp \hat{\mathbf{n}}_0$: (a) out-of-plane configuration formed by transparent indium tin oxide (ITO) electrodes deposited onto the glass plates with separation between them being 10-60 µm as determined by spacers fixing the thickness of the LC slab; (b) in-plane configuration formed by two aluminum strip electrodes separated by 10-15 mm gap. Since $\epsilon_a < 0$, there is no director reorientations far away from the sphere. The in-plane geometry in Fig. 1(b) will be used to explore the two different nonlinear mechanisms of electrophoretic propulsion in the LC. The out-of plane geometry in Fig. 1(a) is used to demonstrate the ability of LCEEP to produce forces that can shift the particles by overcoming other forces, such as gravity and elastic interaction between the director distortions around the particles and the bounding walls.

## III. EXPERIMENTAL RESULTS

*(a) Brownian motion.* If there is no voltage, the spheres in the nematic host experience Brownian motion controlled by two different self-diffusion coefficients $D_\parallel$ and $D_\perp$, associated with the motion parallel and perpendicular to $\hat{\mathbf{n}}_0$, respectively. Loudet et al. described Brownian motion of spheres with quadrupolar director distortions [25]. In our case of perpendicular surface anchoring and dipolar distortions, presence of the topological defect near the sphere breaks the "*fore* − *aft*" symmetry. However, this feature alone does *not* rectify Brownian diffusion of the sphere and does not result in unidirectional movement, Fig. 2. The time dependence of the mean square displacements (MSD) along and perpendicular to $\hat{\mathbf{n}}_0$ follows the classic linear law, Fig. 2(a), at least for the time scales larger than 100 ms, with an anisotropic ratio $D_\parallel / D_\perp = 1.7$. As these times are larger than the director relaxation time, the possible influence of director fluctuations on the linear time dependence of MSD vanishes. In the absence of **E**, the time-average displacement of the particles averages to zero, as clear from the probability distribution $P$ of particle displacements $\delta$ parallel and perpendicular to $\hat{\mathbf{n}}_0$, Fig. 2(b).

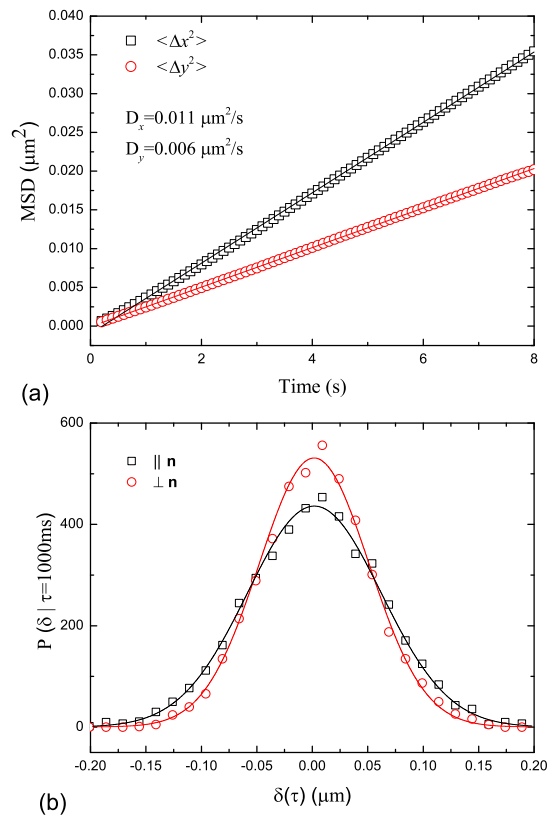

Figure 2: (a) Mean square displacement vs time lag of $2a = 5.08$ µm silica spheres dispersed in E7 LC in the directions parallel ($y$) and perpendicular ($x$) to the director $\hat{\mathbf{n}}_0$ measured at $T = 25$ °C; solid lines represent linear fits. (b) Probability distribution of particle displacement along and perpendicular to the director for time intervals of 1s; solid lines represent Gaussian fits.

*(b) Dielectric reshaping of the director field near the sphere.* In Fig. 1, the electric field does not perturb the overall director, since $\epsilon_a < 0$. Near the sphere, however, the director deviates from $\hat{\mathbf{n}}_0$, and is generally not perpendicular to **E**. The resulting dielectric torque modifies the director around the sphere, shrinking the region of distortions associated with the hyperbolic hedgehog. In principle, this field-induced dielectrically-mediated director reorientation can contribute to the propulsion of particle. To eliminate this dielectric mechanism of propulsion, in our previous work [21], we have performed an experiment with a dielectrically compensated mixture $\epsilon_a = 0$, and found that the particles still move with velocity $v \propto E^2$. The result indicates that besides the dielectric reshaping of the director there are other, more general, driving mechanisms of LC electrophoresis, for example, asymmetric ionic currents around the sphere. In the next two sections, we describe electrophoresis in a LC for two geometries depicted in Fig.1.

*(c) Electrophoresis in out-of-plane geometry.* We tracked the position of DDMAC-coated spheres of diameter $2a = (9.6 \pm 1)\mu$m in a well aligned LC, $\hat{\mathbf{n}}_0 = (0,1,0)$, as a function of a vertical DC field $E_z$, Fig 1(a), by using fluorescent confocal polarizing microscopy (FCPM)[26]. The nematic LC was doped with a very small amount (0.01 wt%) of fluorescent dye N,N-Bis(2,5-di-tert-butylphenyl)-3,4,9,10-perylenedicarboximide (BTBP) and the mixture was sandwiched between two ITO-coated glass plates with a gap $h = 50\mu$m between them. The intensity of fluorescence is maximum when the transition dipole of the dye molecules (parallel to local $\hat{\mathbf{n}}$) is parallel to the polarization of probing light [26]. The silica particles appear dark, whereas the hyperbolic hedgehog appears as a bright spot next to the sphere, Fig.3(b),(c). The LC environment keeps the colloidal particles in the state of levitation [19], since the director distortions around the particles are elastically repelled from the bounding substrates. The effect facilitates the study of dynamic phenomena as it overcomes gravity forces and keeps the particles away from the bounding plates. The vertical electric field $E_z$ displaces the particles along the vertical $z$-axis, Fig.3, and also along the horizontal $y$-axis, Fig. 4. Figure 3(a) shows the field dependence of vertical displacement (measured between the bottom substrate and the center of sphere). There are two mechanisms contributing to this vertical displacement. First, the electric field modifies the director near the spheres and thus alters the elastic forces of repulsion from the bounding walls [27, 28]. This dielectric effect does not depend on the field polarity. The second effect is the polarity-dependent electrophoretic shift, Fig. 3(a). The DDMAC-treated spheres move towards the anode, suggesting that their electric charge is positive. The horizontal motion of particles, caused by the vertical DC field, Fig.1(a), is directed along $\hat{\mathbf{n}}_0$ and $\mathbf{p}$, i.e. the sphere leads the motion. The spheres with $p_y > 0$ and $p_y < 0$ move in antiparallel directions. When the field polarity is reversed, the particles continue to move *in the same direction*, dictated by $\mathbf{p}$. This feature indicates that the electrophoretic velocity $\mathbf{v}_y$ does not depend on the surface charge. Furthermore, it immediately offers an opportunity to drive the particles by an AC field. Figure 4 shows how the velocity $v_y$ depends on the frequency $f$ of the applied electric field for cells of different thickness and with different directions of $\mathbf{p}$. The dependence $v_y(f)$ is non-monotonous with a maximum at $f \approx (0.1-0.5)$Hz. At low frequencies, $\mathbf{v}_y$ is parallel to $\mathbf{p}$. As the frequency increases to $f \approx (10-50)$Hz, the particles reverse the direction of motion, so that $\mathbf{v}_y$ and $\mathbf{p}$ are antiparallel to each other. However, the motion in the reversed direction is slow, Fig. 4, and eventually vanishes at $f > 100$Hz.

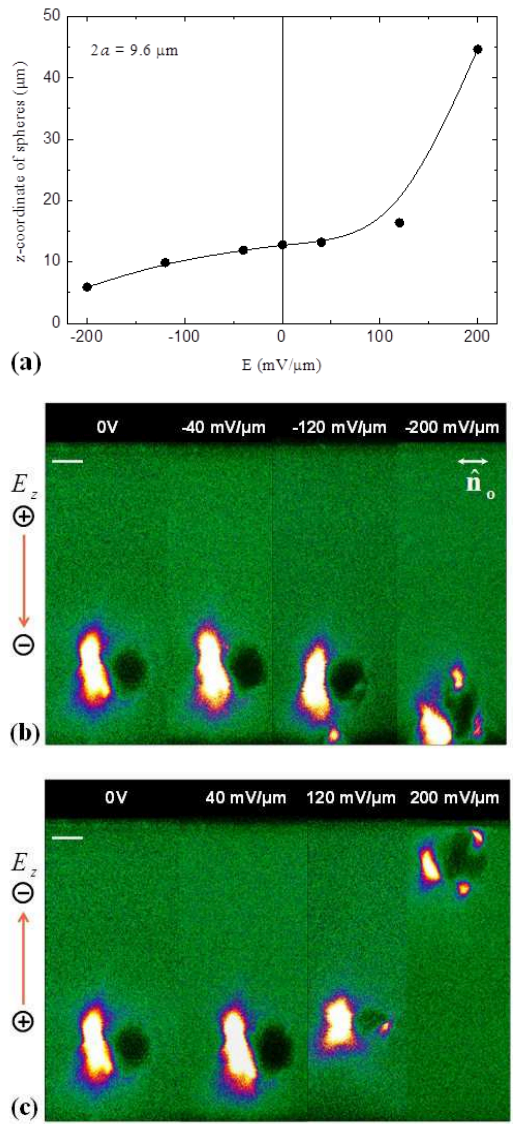

Figure 3: Electrically controlled levitation of $2a = 9.6\mu$m spherical particles in MLC7026-000. (a) $z$-position of microspheres in the bulk as a function of the electric field. FCPM textures of a vertical cross-section of the nematic cell with a colloidal sphere, for (b) negative and (c) positive $E_z$. The white scale bar corresponds to $5\mu$m.

We noticed a small pretilt effect in thick samples: particles with $\mathbf{p}$ parallel to the rubbing direction at the closest (bottom) substrate moved at a slightly higher electrophoretic velocity than their counterparts with an opposite $\mathbf{p}$. In thick samples, the levitating spheres of mass density higher than that of the LC, are located closer to the bottom plate [19]. The difference in elastic repulsion from the wall experienced by the tilted elastic dipoles might be the reason for the different electrophoretic mobilities of spheres with the opposite $\mathbf{p}$s. The difference vanishes when the gap between substrates is reduced (see the data for $h = 11\mu$m in Fig. 4). In thin cells, the particles are close to the middle plane (since the elastic repulsion from both plates is much stronger than the gravity force) [19, 28], and the competing effects of the rubbing directions at the opposite plates cancel each other. The experiments suggest that there are two different mechanisms of electrophoresis in a LC, one related to the electric surface charge of the particles and causing their motion along the applied field, and another one independent of the charge, causing the particles to move in the direction perpendicular to $\mathbf{E}$. In order to explore these mechanisms in a greater detail, we use a different geometry with the in-plane electric field $E_x$, Fig. 1(b). The advantage is that the electrophoretic forces are not competing with gravity and elasticity-mediated forces; the in-plane field $E_x$ causes electrophoretic motion in the horizontal $xy$ plane of the cell.

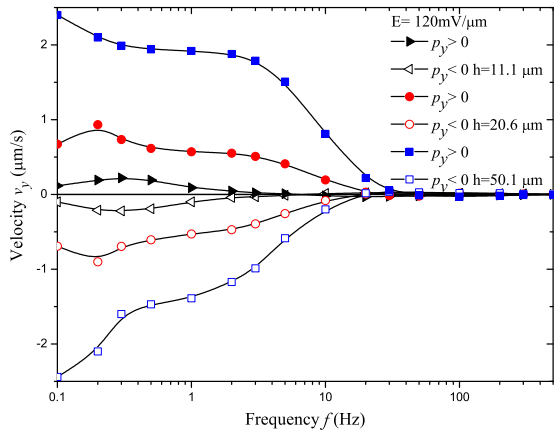

Figure 4: Velocity ($y$-component) of DDMAC-coated spherical particles with $2a = 5.08\mu$m in MLC7026-000 vs frequency of the sinusoidal AC field; out-of plane geometry, $\mathbf{E} = (0, 0, E_z)$ and $\hat{\mathbf{n}}_0 = (0, 1, 0)$. The solid lines are guides to the eye.

*(d) Electrophoresis in in-plane geometry.* The field $\mathbf{E} = (E_x, 0, 0)$ was applied using two parallel aluminum strip electrodes separated by a gap $L = 10 - 15$mm. For quantitative characterization of electrophoresis, one should take into account the finite times of electric charging in the system. Since LCs always contain some amount of ions (caused, e.g., by degradation of molecules), impurities from the alignment layers, etc. [29], the voltage profile across the cell is time-dependent. Once the DC field is applied, the ions move towards the electrodes and build electric double layers near them (screening the applied field), within a characteristic time $\tau_e = \lambda_D L/2D$ [30, 31]. For a typical Debye screening length in LC $\lambda_D = 0.5\mu$m [32], $L = 10$mm, and ionic diffusion coefficient $D = 3 \times 10^{-11}$ m$^2$/s [33], one estimates $\tau_e = 10^2$s . In the DC field experiment, to avoid the electrode screening effects, we measured the particles velocities within a fixed time interval $\Delta t$, equal about 3 min. at low fields ($| E | < 10$mV/$\mu$m) and 1 min. or less at high fields ($| E | > 10$mV/$\mu$m). We experimentally established that within the specified values of $\Delta t$, the spheres moved with the time-independent velocity. After the velocity was measured for a given amplitude and polarity of $\mathbf{E}$, the polarity of the field was reversed and the measurement was repeated, within the same time interval $\Delta t$. The horizontal DC field $\mathbf{E} = (E_x, 0, 0)$ causes the particles to move along the $x$ and $y$ axes, Fig. 5. Consider first the velocity component $v_x$ that is parallel to the applied field $E_x$. The dependence $v_x(E_x)$ is practically linear for small fields, similarly to the electrophoresis in regular fluids, described by the Smoluchowskis formula:

$$\mathbf{v} = \mu_1 \mathbf{E}, \tag{1}$$

where $\mu_1 = \epsilon_m \zeta/\eta$ is the electrophoretic mobility, $\zeta$ is the zeta potential of the particle, $\epsilon_m$ and $\eta$ are the dielectric permittivity and the effective viscosity of the medium, respectively. For stronger fields, one clearly observes a non-linear behavior with a cubic term:

$$v_{13} \equiv v_x = \mu_1 E_x + \mu_3 E_x^3, \tag{2}$$

where $\mu_3$ is a field-independent coefficient. The effect is stronger for larger particles, Fig. 5. The cubic nonlinearity for a nematic LC was also reported by Ryzhkova et al [3].

The velocity component in the direction perpendicular to the electric field shows an absolutely different quadratic behavior:

$$v_{LCEEP} \equiv v_y = \beta E_x^2, \tag{3}$$

where $\beta$ is the field-independent coefficient. The sign of $\beta$ changes with the reversal of the elastic dipole $\mathbf{p}$. The latter feature indicates that the electrophoretic velocity $\mathbf{v}_y$ perpendicular to $\mathbf{E}$ is not related to the net charge of the particles and that the mechanism of this quadratic electrophoresis is rooted in the dipolar distortions of LC around the particle. Once the LC is heated into an isotropic melt, the quadratic effect disappears, $v_{LCEEP} \equiv v_y = 0$, while $v_{13} \equiv v_x$ remains nonzero.

The experiments above suggest that in the nematic LC, the electrophoretic velocity-field relationship $\mathbf{v}(\mathbf{E})$ is tensorial,

$$v_i = \mu_{1ij} E_j + \beta_{ijk} E_j E_k + \mu_{3ijkl} E_j E_k E_l, \tag{4}$$

with tensorial coefficients $\beta_{ijk}$ (rather than a single constant $\beta$) that relate the velocity components $v_i$ to the field components $E_j(i, j = x, y, z)$. Using the experimental data in Fig. 5 and Eq. (4), we find that for DDMAC-coated positively charged glass spheres of diameter $2a = 5.08\,\mu\text{m}$ in the LC MLC7026-000, $\mu_{1xx} = 0.011\,\mu\text{m}^2/\text{mV·s}$, $\beta_{yxx} = 1.5 \times 10^{-4}\,\mu\text{m}^3/\text{mV}^2\text{·s}$ and $\mu_{3xxxx} = 5.2 \times 10^{-6}\,\mu\text{m}^4/\text{mV}^3\text{·s}$. For larger spheres of the same type, $2a = 17.3\,\mu\text{m}$, we find $\mu_{1xx} = 0.038\,\mu\text{m}^2/\text{mV·s}$, $\beta_{yxx} = 7.4 \times 10^{-4}\,\mu\text{m}^3/\text{mV}^2\text{·s}$ and $\mu_{3xxxx} = 1.1 \times 10^{-4}\,\mu\text{m}^4/\text{mV}^3\text{·s}$. The values of $\mu_1$ and $\beta$ are close to those reported previously for positively charged borosilicate spheres of diameter $2a = 17.3\,\mu\text{m}$ placed in the LC mixture E7 with positive dielectric anisotropy ($\mu_{1yy} = 0.07\,\mu\text{m}^2/\text{mV·s}$, $\beta_{yyy} = 5 \times 10^{-3}\,\mu\text{m}^3/\text{mV}^2\text{·s}$) [21]. The data above show that $\mu_{1xx}$ tends to increase with the radius $a$ of the spheres. The result might be related to two effects. First, within the classical theory of electrophoresis in an isotropic fluid, the mobility is expected to grow with the size of the particle when $a$ is much larger than the Debye screening length $\lambda_D$ [34]. The typical values of $\lambda_D$ in LCs are less than 1 $\mu$m, thus we expect $a/\lambda_D > 1$. The second reason is LC-enabled levitation. As shown experimentally by Pishnyak et al [35], larger glass particles levitate closer to the midplane of the cell than the smaller particles (since the elastic forces grow with $a$ faster than the gravity force). Since the viscous drag on the sphere depends on the distance to the bounding plates, the large sphere might experience a weaker effective viscous drag than the small spheres. Note that in our experiment only some of the possible mobility coefficients in Eq.(4) were probed. The main reason is that the finite dielectric anisotropy of the LC restricts the number of configurations in which the field does not perturb the director. By extending the studies to the LC with $\epsilon_a = 0$, one can explore a much wider set of mobility coefficients. A LC with $\epsilon_a = 0$ can be prepared by mixing two LC materials with $\epsilon_a < 0$ and $\epsilon_a > 0$ in an appropriate proportion, or by using a so-called dual-frequency material, in which $\epsilon_a$ changes its sign at a particular frequency of the applied electric field.

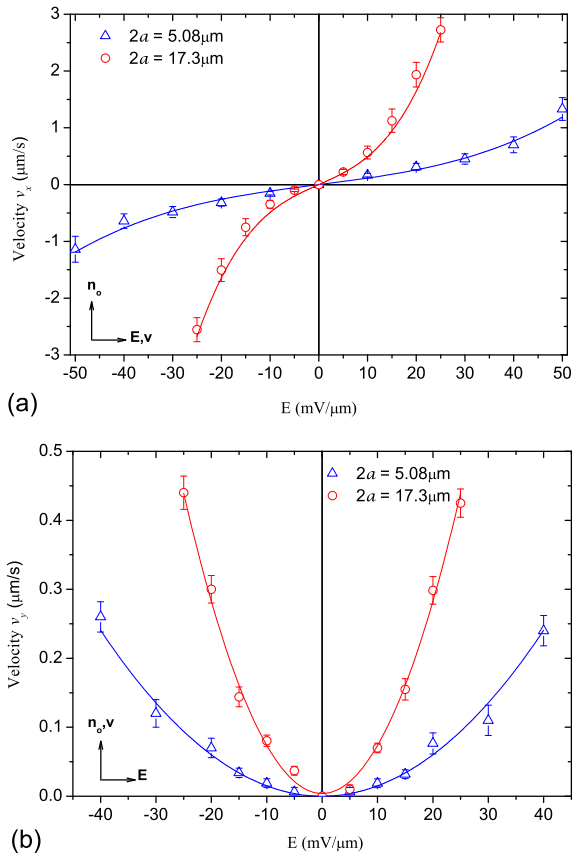

Figure 5: Electrophoretic velocity of positively charged DDMAC-coated dielectric spheres of diameter $2a = 5.08\,\mu\text{m}$ and $2a = 17.3\,\mu\text{m}$ with $p_y > 0$ in the nematic phase ($25\ ^\circ\text{C}$) of MLC7026-000 as a function of the uniform DC electric field; in-plane geometry with $\mathbf{E} = (E_x, 0, 0)$ and $\hat{\mathbf{n}}_0 = (0, 1, 0)$; (a) $x$-component of the velocity, parallel to $\mathbf{E} = (E_x, 0, 0)$ and perpendicular to $\hat{\mathbf{n}}_0 = (0, 1, 0)$, showing a cubic dependence on $E$; (b) $y$-component of the velocity, parallel to $\hat{\mathbf{n}}_0 = (0, 1, 0)$, showing a quadratic dependence on $E$. Solid lines represent the fitting of experimental data using Eq. (2) and Eq. (3) for parts (a) and (b), respectively.

## IV. DISCUSSION

Our experiments demonstrate the existence of two types of electrophoretic motion in a LC of negative dielectric anisotropy, with two very different velocity-field dependences, namely, (1) velocity is an odd function of $E$, $v_{13} \propto E + \alpha E^3$ and (2) an even function of $E$, $v_{LCEEP} \propto E^2$. The first effect is known for isotropic fluids as Stotz-Wein effect [22, 36]. The second effect is observed only if the material is in the LC state; this is why we call it an LC-enabled electrophoresis (LCEEP). Below we discuss both effects for the geometry of Fig.1(b), in which the experimental velocities follow Eq.(2) for the LC analog of Stotz-Wein effect and Eq.(3) for the genuine LCEEP, as schematically summarized in Fig.6.

*(a) Stotz-Wein effect in a LC.* In the experiment, the velocity $v_{13} \equiv v_x$, Eq.(2) and Fig. 5(a), follows a classical linear dependence on the field, when the field is weak. The direction of particle motion changes to the opposite when the field polarity is reversed. This behavior indicates that the electrophoretic mobility is determined by the surface charge of the particle. At higher fields, the cubic term in dependency becomes apparent, Fig. 5(a). In the isotropic fluids, this Stotz-Wien effect [22, 36] is explained by the field-induced polarization of the electric double layers that shifts the electric potential at the particle surface by a quantity $\propto E^2$, so that $v_{13} = \mu_1 E + \mu_3 E^3$ [37]. A similar effect is natural for a LC medium, where it can also be enhanced by the director realignment in the field caused by the dielectric anisotropy, with the realigning torques $\propto E^2$.

The field-induced change $\propto E^2$ in the mobility $\mu_1$ is equivalent to the appearance of the nonlinear coefficient $\mu_3$ in Eq.(2).

*(b) Liquid crystal-enabled electrophoresis.* The LCEEP driven by AC and DC electric fields is observed for spherical particles when the director distortions around them are of dipolar type, Fig. 6. If the particles have a quadrupolar symmetry, LCEEP vanishes [21]. These results indicate that the mechanism for LCEEP is rooted in the type of director distortions which violate the fore-aft (left-right) symmetry of the LC medium around the sphere. As already mentioned, anisotropy of both dielectric and conductive properties of the nematic LC can contribute to the propulsion of the particles in the electric field. The dielectric anisotropy of the LC causes director realignment in the applied electric field that is especially pronounced in the region of the hedgehog (as compared to the opposite side of the sphere where the director is more homogeneous). Although further exploration of this effect is important, we note that if the dielectric anisotropy is reduced to zero, the LCEEP still persists [21], suggesting that the dielectric anisotropy is not a necessary condition for LCEEP. A more general mechanism is associated with the different mobility of electric ions around the sphere with the dipolar director distortions [21]. Consider a dielectric sphere with a positive surface charge, screened by a cloud of negative counterions, Fig.6. Besides these two types of charges, the LC bulk also contains mobile ions of both positive and negative charge (the system as a whole is electrically neutral). Once the electric field is applied, the mobile ions in the LC start to move in two opposite directions, from the two poles of the sphere towards its equator. The flow pattern (from the poles towards the equator) does not change with the field reversal. Since the mobility of ions is anisotropic in a LC (mobility along the director is typically higher than in the perpendicular direction), the velocity of ions in the hyperbolic hedgehog region is expected to be different from the velocity at the opposite, hedgehog-free side of the sphere, resulting in a broken mirror symmetry with respect to the equatorial plane of the sphere that is perpendicular to the elastic dipole **p**. Furthermore, since the viscosity of the LC medium is also anisotropic, the LC flow triggered around the sphere by ions is also expected to follow a broken symmetry pattern, Fig.6.

These asymmetries give rise to an electrophoretic velocity that is polarity-independent, $v_{LCEEP} \equiv v_y \propto E_x^2$. For particles moving perpendicular to the electric field, the director field asymmetry is the only reason for electrophoretic propulsion since $v_y$ is independent of the particle charge. Of course, if the particle is charged, then there is also a regular electrophoretic effect shown in Fig. 5(a) with the velocity $v_{13} \equiv v_x$ parallel (or antiparallel, depending on the sign of charge) to the direction of the applied field $E_x$, as in Eq. (2). For uncharged particles $v_x = 0$, and only the quadratic contribution to LCEEP would survive. If the driving field is of an AC type, then the average value of $v_{13} \equiv v_x$ vanishes but the component $v_{LCEEP} \propto E_x^2$ remains non-zero, allowing an AC control of LCEEP. The frequency dependence of electrophoretic velocity has been extensively studied for ICEP in isotropic fluids [4, 9, 38] and recently by us for LCEEP [21]. In both cases, there are two important time scales: 1) a characteristic charging time $\tau_c = \epsilon_m \lambda_D^2 / \epsilon_d D$ of a dielectric sphere; 2) the characteristic electrode charging time $\tau_e = \lambda_D L / 2D$. If only these two scales are relevant, and if the Debye layer around the sphere is thin and uniform, the velocity shows a frequency dependence

$$v(\omega) = v_o \frac{\omega^2 \tau_e^2}{(1+\omega^2 \tau_c^2)(1+\omega^2 \tau_e^2)},$$

with a single peak at the characteristic charging frequency $\omega_c = \tau_c^{-1}$ of the electric double layer. This simplified model cannot capture all the features of the frequency dependence of velocity presented in Fig. 4 which demonstrates two peaks. The model can be improved by taking into account that the cloud of counterions around the spherical particle in a LC matrix is not spherically symmetric; this is a subject of further studies. To conclude, we demonstrated a rich variety of nonlinear electrophoretic effects when the moving particle is located in a LC rather than in an isotropic fluid. The electric field is applied in such a way that it does not distort the director far away from the particle. The electric field causes electrophoresis with velocity components that are parallel and orthogonal to the field. In the first case, the velocity is proportional to the linear and cubic power of the field. In the second case, the velocity is quadratic in the field, demonstrating a new mechanism of electrophoresis that we call a *liquid crystal-enabled electrophoresis*, or LCEEP. The LCEEP occurs when the director distortions around the inclusion show dipolar symmetry. The LCEEP effect makes it possible to move the particles by an AC electric field even when the field is absolutely balanced, i.e., not only when its time average is zero $\langle E \rangle = 0$, but also when the time average of higher moments are zero, e.g., $\langle E^3 \rangle = 0$. This represents an important difference with the so-called aperiodic nonlinear electrophoresis in isotropic fluids with velocity that is an odd function of the electric field [39]. It also allows one to transport particles of zero charge and particles of an absolutely symmetric shape and physical properties. The experiments presented above clearly demonstrate the tensorial relationship between the electrophoretic velocity in the LC and the electric field. The fact that the particles move perpendicular to **E** in LCEEP means that the driving voltage can be applied across the thickness of the channel (few μm) rather than along the pathway (several cm) in electrophoretic devices; this effect allows one to use modest voltage sources to achieve high fields, which might be especially beneficial for portable devices. The possibility of moving the particle in different directions without altering the direction of the field is also remarkable. The field and frequency dependencies shown in Fig. 4 and 5 indicate that one can combine two differently oriented driving fields with different frequencies to control the overall 3D trajectory of particles. Further diversification can be achieved by using LCs with pre-patterned director field and by using LCs of zero dielectric anisotropy $\epsilon_a = 0$ so that the electric field does not cause the director reorientation regardless of the mutual orientation of $\hat{\mathbf{n}}_0$ and **E**.

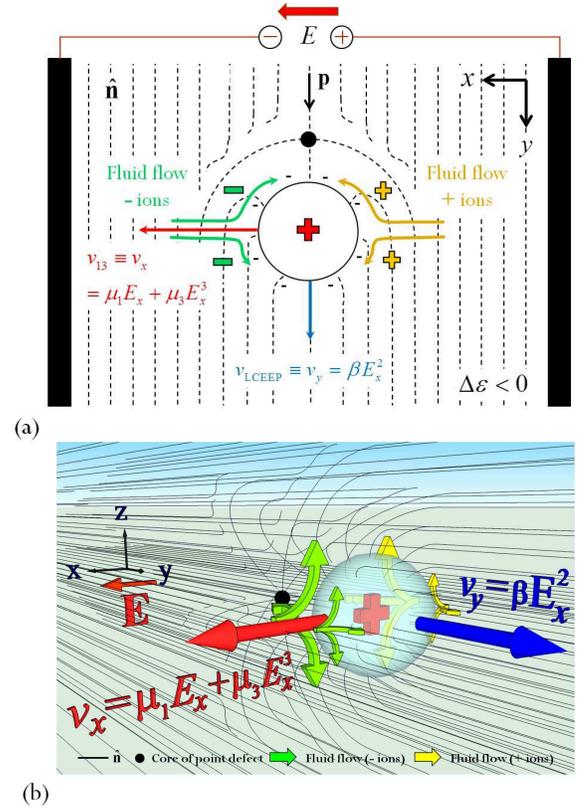

(a)

(b)

Figure 6: Schematic of two types of electrophoretic motion of a sphere with normal anchoring at the surface, in a uniformly aligned nematic LC: (a) in-plane $(xy)$ two-dimensional representation; (b) three-dimensional view.

*The authors gratefully acknowledge discussion with S. Klein. T. Lubensky, S. V. Shiyanovskii and thank L. Tortora for help with surface functionalization of particles. The research was supported by NSF DMR 1104850.*